\documentclass[conference]{IEEEtran}
\makeatletter
\def\ps@headings{%
\def\@oddhead{\mbox{}\scriptsize\rightmark \hfil \thepage}%
\def\@evenhead{\scriptsize\thepage \hfil \leftmark\mbox{}}%
\def\@oddfoot{}%
\def\@evenfoot{}}
\makeatother
\usepackage[letterpaper, left=0.625in, right=0.625in, bottom=1in, top=0.75in]{geometry}
\usepackage[pdftex]{graphicx}
\usepackage{cite}
\usepackage{amsmath}
\usepackage{mathtools}
\usepackage{enumerate}
\usepackage{adjustbox}
\usepackage{algorithm}
\usepackage{algpseudocode}
\usepackage[caption=false,font=footnotesize]{subfig}
\newcommand{\doi}[1]{\textsc{doi}: \href{http://dx.doi.org/#1}{\nolinkurl{#1}}}
\usepackage{stfloats}
\usepackage{footnote}
\usepackage[inline]{enumitem}
\usepackage{microtype}
\usepackage{acronym}
\usepackage{xcolor}
\usepackage{colortbl,hhline}
\usepackage{tabularx}
\usepackage{boldline}
\usepackage{multirow,tabu, booktabs}
\usepackage{tikz,pgfplots}
\usetikzlibrary{arrows}
\usepackage{pgfplotstable}
\usepgfplotslibrary{units,groupplots}
\usepackage{pifont}
\usepackage{amssymb}
\usepackage{amsfonts}
\usepackage{threeparttable}
\usepackage{fancyhdr}

\newcommand{\highlightnewtext}{1}

\ifnum\highlightnewtext=0

\newcommand{\edited}[1]{{\leavevmode\color{orange}{#1}}}

\else

\newcommand{\edited}[1]{{\leavevmode\color{black}{#1}}}

\fi

\newcommand{\keepcomment}{1} 

\ifnum\keepcomment=1
	\newcommand{\roberto}[1]{{\leavevmode\color{blue}{[ROB: #1]}}}
	\newcommand{\domenico}[1]{{\leavevmode\color{magenta}{[DOM: #1]}}}
	\newcommand{\maged}[1]{{\leavevmode\color{orange}{[MAG: #1]}}}
	\newcommand{\sandra}[1]{{\leavevmode\color{green}{[SSH: #1]}}}
\else
	\newcommand{\roberto}[1]{\ignorespaces\unskip}
	\newcommand{\domenico}[1]{\ignorespaces\unskip}
	\newcommand{\maged}[1]{\ignorespaces\unskip}
	\newcommand{\sandra}[1]{\ignorespaces\unskip}
\fi

\newlength{\Oldarrayrulewidth}

\pgfplotsset{
	compat = newest,
	tick label style={font=\sffamily\tiny},
	label style={font=\sffamily\tiny},
	legend style={font=\sffamily\small\raggedleft, mark options={mark size=1.5pt}},
	legend cell align=left,
	legend image code/.code={
                            \draw[mark repeat=2,mark phase=2]
                            plot coordinates {
                            (0cm,0cm)
                            (0.2cm,0cm)        
                            (0.4cm,0cm)         
                            };%
                            },
	grid style={dotted,gray}
}

\newcommand{\cmark}{\ding{51}}%
\newcommand{\xmark}{\ding{55}}%

\graphicspath{{../}{graphs/}}
\makeatletter
\renewcommand{\fnum@figure}{Figure \thefigure}
\makeatother
\begin{document}
\newcommand{\ourtool}[0]{\text{GADoT}}

\newcommand*{\Scale}[2][4]{\scalebox{#1}{#2}}%

\title{GADoT: GAN-based Adversarial Training \\ for Robust DDoS Attack Detection}


\author{
	Maged Abdelaty$^{1,3}$,
	Sandra Scott-Hayward $^2$,
	Roberto Doriguzzi-Corin$^1$, and 
	Domenico Siracusa$^1$\\
	
	\small{$^1$ICT, Fondazione Bruno Kessler - Italy} \quad
	\small{$^2$CSIT, Queen’s University Belfast - Northern Ireland} \\
	\small{$^3$University of Trento - Italy} \\
	\small{Email: mabdelaty@fbk.eu, s.scott-hayward@qub.ac.uk, rdoriguzzi@fbk.eu, dsiracusa@fbk.eu}
}

\maketitle              

\thispagestyle{fancy}
\renewcommand{\headrulewidth}{0pt}
\chead{\scriptsize This is the author's version of an article presented at the 2021 IEEE Conference on Communications and Network Security. \\Changes were made to this version by the publisher prior to publication.}
\cfoot{\scriptsize Copyright (c) 2021 IEEE. Personal use is permitted. For any other purposes, permission must be obtained from the IEEE by emailing pubs-permissions@ieee.org.}

\acrodef{cnn}[CNN]{Convolutional Neural Network}
\acrodef{rnn}[RNN]{Recurrent Neural Network}
\acrodef{fnn}[FNN]{Feedforward Neural Networks}
\acrodef{snn}[SNN]{Self-normalizing Neural Network}
\acrodef{svm}[SVM]{Support Vector Machine}
\acrodef{gmm}[GMM]{Gaussian Mixture Model}
\acrodef{knn}[KNN]{K-Nearest Neighbour}
\acrodef{ddos}[DDoS]{Distributed Denial of Service}
\acrodef{aml}[AML]{Adversarial Machine Learning}
\acrodef{ml}[ML]{Machine Learning}
\acrodef{gan}[GAN]{Generative Adversarial Network}
\acrodef{nids}[NIDS]{Network Intrusion Detection System}
\acrodef{ids}[IDS]{Intrusion Detection System}
\acrodef{wgan-gp}[WGAN-GP]{Wasserstein GAN with gradient penalty}
\acrodef{wgan}[WGAN]{Wasserstein GAN}
\acrodef{mtu}[MTU]{maximum transmission unit}
\acrodef{tpr}[TPR]{true positive rate}
\acrodef{fpr}[FPR]{false positive rate}
\acrodef{fnr}[FNR]{false negative rate}
\acrodef{lucid}[LUCID]{Lightweight, Usable CNN in DDoS Detection}
\acrodef{sdn}[SDN]{Software-Defined Networks}
\acrodef{iot}[IoT]{Internet of Things}
\acrodef{fgsm}[FGSM]{Fast Gradient Sign Method}
\acrodef{pgd}[PGD]{Projected Gradient Descent}
\acrodef{bfp}[BFP]{Benign Feature Perturbation}
\acrodef{gadot}[GADoT]{GAN-based Adversarial training for robust DDoS attack deTection}
\acrodef{rdpro}[RDPro]{Robust DDoS Protection}
\acrodef{prodm}[ProDM]{Protecting DDoS detection Models against adversarial attacks}
\acrodef{iid}[i.i.d.]{independent and identically distributed}
\acrodef{ood}[o.o.d.]{out-of-distribution}
\acrodef{ai}[AI]{Artificial Intelligence}

\begin{abstract}	
\ac{ml} has proven to be effective in many application domains. However, \ac{ml} methods can be vulnerable to adversarial attacks, in which an attacker tries to fool the classification/prediction mechanism by crafting the input data. In the case of \ac{ml}-based \acp{nids}, the attacker might use their knowledge of the intrusion detection logic to generate malicious traffic that remains undetected.
One way to solve this issue is to adopt adversarial training, in which the training set is augmented with adversarial traffic samples.
This paper presents an adversarial training approach called \ourtool, which leverages a \ac{gan} to generate adversarial \acs{ddos} samples for training. We show that a state-of-the-art \ac{nids} with high accuracy on popular datasets can experience more than $60\%$ undetected malicious flows under adversarial attacks. We then demonstrate how this score drops to $1.8\%$ or less after adversarial training using \ourtool.

\end{abstract}


\section{Introduction}\label{sec:introduction}

The complexity and volume of \ac{ddos} attacks continues to increase, causing significant disruption to critical online services and leading to economic losses.
For instance, the number of \ac{ddos} attacks is expected to reach 15.4 million globally by 2023 \cite{ciscoddos}. Meanwhile, a new record of 2.3\,Tbps was reached in an attack targeting Amazon Web Services (AWS) in Q1 2020 \cite{a10ddos}. To achieve this scale and damage, attackers develop more sophisticated attacks combining multiple vectors (i.e. protocols) and adapting their attacks to bypass \acp{nids}.

Over recent years, \acp{nids} have evolved to leverage \ac{ml} models to cope with the detection of new attacks, which cannot be identified by signature-based detection methods. 
An \ac{ml}-based \ac{nids} essentially consists of a feature extractor and a \ac{ml}-based model.
The feature extractor processes the raw network traffic to derive features that are arranged into data samples suitable to input to the model.
The model is trained using malicious and benign samples and then deployed to classify new samples extracted from live traffic. Several \ac{ml}-based \acp{nids} have been proposed demonstrating detection of \ac{ddos} attacks with high accuracy \cite{mirsky2018kitsune, doriguzzi2020lucid, elsayed2020ddosnet}.

However, in the same way that an attacker might adapt their attack to remain just below the threshold of a threshold-based detection system or modify their attack to avoid signature-based detection, \ac{ml}-based \acp{nids} are vulnerable to adversarial attack. The adversarial attack aims to evade the classification model and thus reach the target, achieving the planned \ac{ddos}. In a \ac{ddos} scenario, an attacker generates perturbed \ac{ddos} flows such that the \ac{nids} extracts samples incorporating features slightly different from those of unperturbed flows. The ability to lead the \ac{ml} model to misclassify adversarial samples and hence reduce the \ac{nids} accuracy has been demonstrated in previous works \cite{kuppa2019black, aiken2019investigating}.

\acfp{gan} are deep generative models with the capability of learning the distribution of the input data, then producing new similar examples.        
A \ac{gan} consists of two models: a discriminator and a generator that are trained to compete with each other.
The generator model captures the training data distribution to produce fake samples resembling those from the training data.
Our intuition here is that \ac{gan}-based sample generators can be exploited to enrich the training set with adversarial examples similar to those that an attacker would use to fool the \ac{nids}, that is, \ac{ddos} examples with the feature distribution of the benign samples.

In this paper, we present \ac{gadot}, an adversarial training framework designed to produce strong adversarial examples, hence enabling an effective adversarial training of \ac{ml}-based \ac{ddos} classifiers.
Our framework employs the generator model of a \ac{gan} trained on benign samples to produce fake-benign samples.
Using these fake-benign samples, we perturb the \ac{ddos} samples by replacing their features with values from those fake samples.
The perturbed samples are added to the training dataset, which already contains benign samples and unperturbed \ac{ddos} samples, creating an augmented dataset for adversarial training.
The trained model is expected to classify \ac{ddos} samples correctly, whether they are perturbed or not.
To the best of our knowledge, \ourtool\ is the first solution to use \ac{gan}-based perturbations that decouples the trained model from the generation process of adversarial samples to prevent the model from influencing the strength of adversarial samples \cite{tramer2017ensemble}.


There are two approaches to implementing adversarial attacks against \ac{nids}; perturbing data samples and perturbing network traffic\footnotemark.
\footnotetext{Formally, this equates to the feature space and the problem space, respectively, as studied in \cite{pierazzi2020intriguing}.}
In the first approach, the adversary introduces perturbations directly on the preprocessed samples before feeding them to the \ac{ml} model.
In the second approach, the adversary generates adversarially perturbed \ac{ddos} flows, which are processed by the feature extractor to create data samples to be classified by the \ac{ml} model.
This approach can be more realistic than perturbing extracted samples that require the adversary to bypass the feature extractor to feed such samples to the model.

For a practical evaluation, we adopt the approach of perturbing the network traffic to generate two datasets in which the semantics of \ac{ddos} attacks are fully preserved.
The first dataset contains six traces with adversarial SYN flood attacks \cite{syn-flood}, while the other is based on the CICIDS2017 dataset \cite{sharafaldin2018toward} and contains traces with adversarial HTTP GET flood attacks \cite{get-flood}. The evaluation results on both datasets are the basis of our conclusions on the performance of \ourtool. To the best of our knowledge, \ourtool\ is the first solution to be evaluated against representative real-world network attack scenarios.


The contributions of this paper are the following:
\begin{itemize}
    \item We propose \ourtool, which is an adversarial training approach to increase the robustness of \ac{ml} models used for \ac{ddos} attack detection.
    The approach exploits a \ac{gan} to generate fake-benign samples in order to perturb \ac{ddos} samples.
    The training dataset is augmented with these perturbed samples in order to train adversarially robust models.
    \item We provide an analysis of single and multiple (combined) feature perturbations to offer insight into the model's vulnerability before and after adversarial training.
    \item We offer a practical evaluation and analysis of the robustness of the \ac{ml} model trained with \ourtool\ by using network traffic traces that capture adversarially perturbed SYN and HTTP \ac{ddos} flood attacks. 
\end{itemize}

The rest of the paper is structured as follows: Section \ref{sec:Related_Work} reviews the related work on adversarial attacks against \acp{nids} and the proposed solutions to increase the robustness of \ac{ml}-based \acp{nids}.
Section \ref{sec:threat_model} introduces the threat model of this work.
The proposed \ourtool\ approach is explained in Section \ref{sec:Approach}.
Section \ref{sec:exp_design} presents the datasets, details of the specific feature perturbation and experiment design.
The results are discussed in Section \ref{sec:results}.
Finally, Section \ref{sec:conclusion} concludes the paper.



\section{Related work} \label{sec:Related_Work}
This section reviews the recent advances in the evaluation of \acp{nids} in an adversarial setting and the approaches for enhancing the robustness of these systems.

\subsection{Adversarial attacks against \acp{nids}}

As highlighted in Section \ref{sec:introduction}, the vulnerability of learning-based \acp{nids} to adversarial perturbation has been explored and implemented by means of two approaches: perturbing data samples and perturbing network traffic.

\subsubsection{Perturbing extracted samples}

The key to the success of an adversarial evasion attack is to maintain the characteristic of the attack. For example, for a SYN Flood attack that sends a volume of SYN packets (i.e. SYN flag set), a feature based on the \textit{flags} of the TCP header should not be manipulated. 
Yan et al. \cite{yan2019automatically} split features in the KDDCup99 dataset into \textit{unchangeable} that are essential to maintain the malicious function and \textit{changeable} than can be manipulated without affecting this function.
Then, the authors employed a \ac{wgan}-based neural network to manipulate the changeable features of the \ac{ddos} attack.
The reported results show a $50 \%$ drop in the detection accuracy of the \ac{nids} on the KDDCup99 dataset.
However, this approach lacks the discussion of how these perturbed samples can be reproduced in live network traffic.
This live traffic should be valid according to the different network protocols in order to pass through the network and carry out the intended attack function on the victim side.

\subsubsection{Perturbing network traces}


Aiken et al. \cite{aiken2019investigating} investigate adversarial attacks targeting an anomaly based \ac{nids} within \ac{sdn}.
The authors propose perturbing three features: payload size, packet rate and volume of bidirectional traffic, which can be implemented in live traffic without compromising the attack function.
The attack is implemented in SYN flood traces that are classified by several \ac{ml} models, including \ac{svm} and \ac{knn}.
The results show that by combining perturbed features, a detection accuracy drop from $100 \%$ to $0 \%$ is observed across some of the classifiers and the \ac{knn} classifier is more robust than the others.

\subsection{Defence mechanisms in \acp{nids}}

In the following, we review some of the related work proposed to increase the robustness of \acp{nids}, which are based on denoising autoencoders \cite{hashemi2020enhancing}, adversarial training \cite{khamis2020evaluation, abou2020investigating} and combining adversarial training with ensemble voting \cite{zhang2020tiki}.

Hashemi et al. \cite{hashemi2020enhancing} propose a solution for increasing the robustness through denoising the samples to remove any adversarial perturbations.
The authors evaluated their approach on the CICIDS2017 dataset and use the adversarial attacks presented in \cite{hashemi2019towards}.
The results suggest that RePO is more robust than other \acp{nids} such as Kitsune \cite{mirsky2018kitsune}.


Zhang et al. \cite{zhang2020tiki} propose a defence approach that combines three techniques: ensemble voting based on three different \ac{ml} models, ensemble adversarial training \cite{tramer2017ensemble}, and query detection.
A flow is classified as benign only if all the models classify it as benign, forcing the adversary to craft samples that can evade all models simultaneously.
Moreover, they use ensemble adversarial training \cite{tramer2017ensemble} to augment the training dataset with adversarial samples, improving the robustness of individual models.
The defence approach is evaluated on the CICIDS2018 dataset \cite{sharafaldin2018toward} and was effectively able to reduce the success rate of the adversarial attacks.

Despite the good performance of these defence approaches, they are only evaluated on perturbations that are implemented on extracted samples, and there is no discussion of how to convert such perturbed samples into real network traffic.
Furthermore, a common limitation of the adversarial training approaches \cite{abou2020investigating, usama2019generative, khamis2020evaluation} is the use of adversarial samples that are generated based on the target/victim models.
In such approaches, the victim model learns to generate weak adversarial samples, instead of learning to defend against strong perturbations and hence remains vulnerable to adversarial attacks \cite{tramer2017ensemble}.

In this work, we propose an adversarial training approach called \ourtool\ in which the target model is decoupled from the generation of the adversarial samples used for training.
We generate the adversarial samples by perturbing \ac{ddos} samples with values obtained from a \ac{gan}.
The trained model thus learns to defend against strong perturbations and can be resistant to adversarial attacks.
Furthermore, we evaluate the proposed approach using realistic network traces that are adversarially perturbed.
These traces capture perturbations that are suitable for volumetric \ac{ddos} attacks.

\section{Threat Model}\label{sec:threat_model}

To characterise the adversary attacking our \ac{ml}-based \ac{nids}, we identify their goals, capabilities and knowledge, as per the threat modeling framework presented by Biggio et al. in \cite{biggio2018wild}.
Our adversary model is summarised in Table \ref{tab:threat_model}.

\begin{table}[t!]
\centering
\caption{Adversary model}
\label{tab:threat_model}
\begin{adjustbox}{width=0.9\linewidth,center}
\begin{tabular}{@{}lll@{}}
\toprule
\textbf{\begin{tabular}[c]{@{}l@{}}Threat Model \\ Characteristic\end{tabular}} & \textbf{Type} & \textbf{\begin{tabular}[c]{@{}c@{}}Adversary \\ View\end{tabular}} \\ \midrule
\multirow{3}{*}{\textbf{Adversary Goals}} & \begin{tabular}[c]{@{}l@{}}Compromising integrity \\ (Evasion)\end{tabular} & \cmark \\ \cmidrule(l){2-3}
 & Compromising availability & \xmark \\ \cmidrule(l){2-3}
 & Compromising confidentiality & \xmark \\ \midrule
\multirow{2}{*}{\textbf{Adversary Knowledge}} & Feature set & \cmark \\ \cmidrule(l){2-3}
 & Satisfy domain constraints & \cmark \\ \midrule
\multirow{4}{*}{\textbf{Adversary Capabilities}} & Manipulate training data & \xmark \\ \cmidrule(l){2-3}
 & Manipulate test data & \cmark \\ \cmidrule(l){2-3}
 & Manipulate model & \xmark \\
 \bottomrule
\end{tabular}
\end{adjustbox}
\end{table}

\subsection{Attacker's goal}
The adversary aims to evade the \ac{ml} model through manipulating the \ac{ddos} traffic to be misclassified as benign. If successful, this operation would allow the attack traffic to remain undetected and to reach the victim machine or network.  
This maps to the online evasion attack as defined in the recently published MITRE adversarial threat matrix \cite{mitre2020adv}.

\subsection{Attacker's knowledge}

The attackers know that the network traffic is monitored by an \ac{ml}-based \ac{nids}.
However, they have no direct access to the \ac{nids} itself or the underlying \ac{ml} model.
As a result, they cannot compromise the \ac{nids} by bypassing some of its components (such as the feature extractor) in order to feed manipulated samples directly to the model. 
Thus, state-of-the-art attacks \cite{abou2020investigating, khamis2020evaluation, yang2018adversarial} based on techniques such as \ac{fgsm}, which manipulate the data samples, are not viable in such an attack scenario.
However, since \ac{ddos} attacks are well researched in the literature, the attackers can exploit their field knowledge to know the basic traffic features employed for attack detection. Based on that, they manipulate these features to evade the classification model.
At the same time, the manipulated traffic must satisfy network domain constraints such that the objective of the attack remains valid. For example, as previously noted, a SYN Flood must be comprised of SYN packets. Similarly, the adversary should ensure that a manipulated packet will not be discarded for reasons such as invalid header fields (e.g., the checksum). These assumptions present a realistic threat model for the \ac{nids} problem space.

\subsection{Attacker's capability}

The attackers can perturb the \ac{ddos} flows at test time in order to evade the classification model, allowing the \ac{ddos} attack to pass undetected.
They can also build upon the published adversarial attacks, such as \cite{aiken2019investigating}, to select the traffic parameters that can be tuned to craft an evasion attack.

\section{The \ourtool\ approach}
\label{sec:Approach}

\begin{figure*}[htp]
	\centering
	\includegraphics[width=0.65\textwidth]{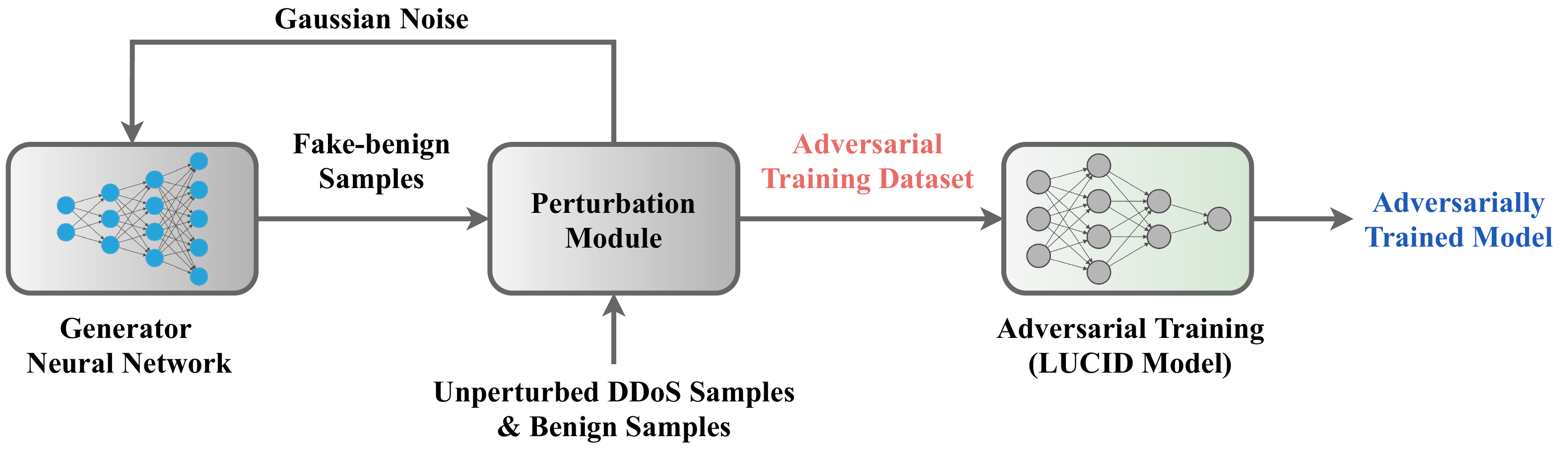}
	\caption{\edited{Illustration of the \ourtool\ approach for adversarial training}}
	\label{fig:aml_defence_territory}
\end{figure*}

To produce a robust deep learning-based \ac{ddos} detector, we propose an adversarial training approach, \ourtool, in which an \ac{ml} model is trained with benign samples, \ac{ddos} samples, and adversarial \ac{ddos} samples.
The goal is to increase the robustness against the perturbed \ac{ddos} samples, in anticipation of an attacker who can craft packets and perturb malicious flows to mimic those characteristics of benign flows.
The approach employs a generator neural network that generates fake-benign samples used to create adversarial samples and a perturbation module that produces the augmented dataset $T_{adv}$ used for adversarial training of the target model, as depicted in Figure \ref{fig:aml_defence_territory}.

In the following, we briefly introduce the target model used in this work, \ac{lucid}\footnotemark \cite{doriguzzi2020lucid}, and how it prepares the data samples before classification, then we explain the modules employed by \ourtool. Finally, we describe two baseline approaches against which \ourtool\ is compared.
\footnotetext{We selected \ac{lucid} as our target model as it achieves the best classification accuracy of the deep learning-based \ac{ddos} detection solutions in the literature. Its preprocessing tool also enables us to adopt the practical approach of perturbing the network traffic.}

\subsection{\ac{lucid} as a target model}
\label{subsec:lucid}

\begin{figure}[h!]
\centering
\includegraphics[width=0.7\columnwidth]{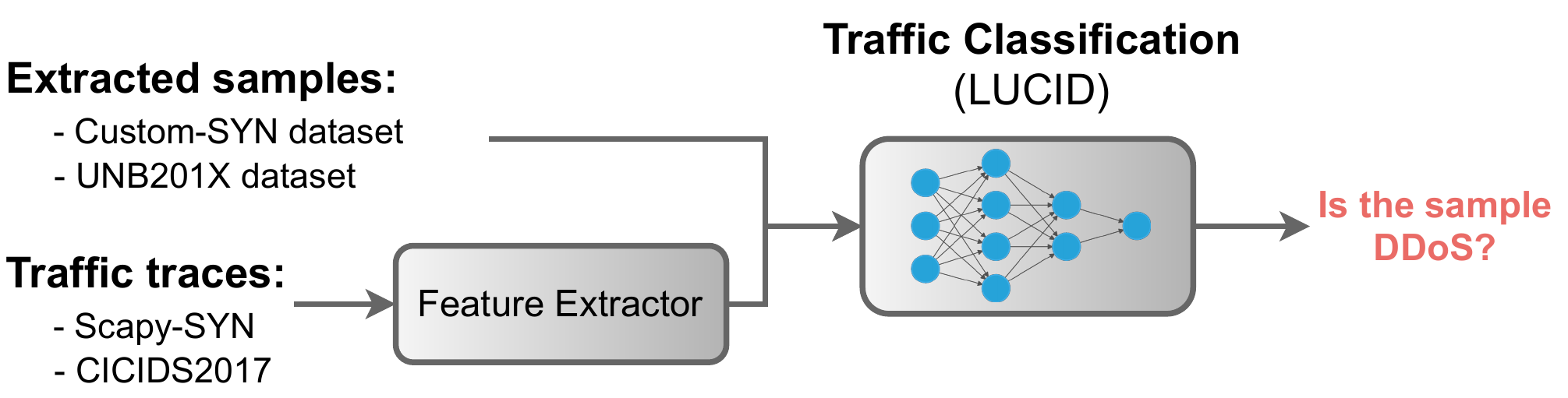}
\caption{Illustration of the traffic classification process}
\label{fig:fext_model}
\end{figure}

\ac{lucid} is an efficient \ac{ddos} detection system that implements a pre-processing tool for the network traffic and a \ac{cnn}. The source code of both pre-processing tool and neural network are publicly available at \cite{lucid-github}.
The preprocessing tool, which we refer to as the feature extractor, converts the flows extracted from the network traffic into data samples compatible with the \ac{cnn}.
In Figure \ref{fig:fext_model}, we illustrate the feature extractor and the classification model.
The figure also contains other details such as dataset names that we introduce later, in Section \ref{sec:exp_design}.
The \ac{cnn} classifies the data samples as either benign or \ac{ddos}.
Each data sample includes a set of features extracted from each packet of a bi-directional flow in a specific time window.

Using the LUCID feature extractor, we extract 11 features that are described in Table \ref{tab:feat_description} and use ten seconds as the time window length. A network flow, or sample, is represented as an array of fixed size 10x11, whose 10 lines are the packets that belong to the flow collected within the time window, while the 11 columns are the features extracted from each packet. The size of the array is fixed and decided at training time, as this is a requirement for the CNN. Flows with more than 10 packets in a time window are truncated, while shorter flows are zero-padded. The number of actual packets in a sample (non zero-padded rows) can be seen as a further feature that we call \textit{Flow length} in the rest of the paper.

\begin{table}[t!]
\centering
\caption{The 11 packet header attributes used in LUCID \cite{doriguzzi2020lucid}}
\label{tab:feat_description}
\begin{adjustbox}{width=0.9\columnwidth,center}
\begin{tabular}{@{}ll@{}}
\toprule
\textbf{Feature} & \textbf{Description} \\ \midrule
Time & \begin{tabular}[c]{@{}l@{}}Relative time of each packet with respect to the \\  first packet in the time window\end{tabular} \\ \midrule
Packet Len & Length of the entire IP packet \\ \midrule
Highest Layer & The highest protocol in a packet, e.g. TCP or HTTP \\ \midrule
IP Flags & \begin{tabular}[c]{@{}l@{}}Three bits taking one of three states: all zeros, \\ do not fragment or more fragments\end{tabular} \\ \midrule
Protocols & Representing the protocols found in the packet \\ \midrule
TCP Len & Size of the TCP segment (header + payload) \\ \midrule
TCP Ack & Relative acknowledgment number \\ \midrule
TCP Flags & Nine 1-bit flags of a TCP segment \\ \midrule
TCP Win Size & \begin{tabular}[c]{@{}l@{}}The value of the field of the same name in a TCP segment\end{tabular} \\ \midrule
UDP Len & Size of the UDP segment (header + payload) \\ \midrule
ICMP Type & The value of the type field of an ICMP message \\ \bottomrule
\end{tabular}
\end{adjustbox}
\end{table}

\subsection{The generator neural network}
\label{subsec:gen_nn}

The generator neural network is the generator model $\mathcal{G}$ of a \ac{wgan-gp} \cite{gulrajani2017improved} that achieves a state of the art performance in terms of training stability and diversity of the generated samples.
In addition to the generator $\mathcal{G}$, the \ac{wgan-gp} includes a discriminator model $\mathcal{D}$ that learns to distinguish between the real samples from the training data and the generator’s output.
During the \ac{gan} training, $\mathcal{G}$ and $\mathcal{D}$ compete with each other.
$\mathcal{G}$ learns to generate better fake samples that fool $\mathcal{D}$, whereas $\mathcal{D}$ learns to label those samples, minimising the success of $\mathcal{G}$.
An optimal generator causes the discriminator to fail to distinguish between the real and generated samples.

Given a training dataset of benign and \ac{ddos} samples, we train the generator and discriminator of the \ac{wgan-gp} with the benign samples.
After training, the discriminator is removed and the trained generator $\mathcal{G}$ of \ac{wgan-gp} becomes the generator of \ourtool\ (Figure \ref{fig:aml_defence_territory}).
This generator takes a source of noise $z$ as an input and generates new samples $\hat{x}=\mathcal{G}(z)$ with the feature distribution of the training data. As \ac{wgan-gp} is trained with benign samples, the resulting $\mathcal{G}(z)$ produces what we call \textit{fake} benign samples.

We chose to generate benign samples in order to perturb the features of the \ac{ddos} samples with values seen in benign traffic.
This matches with the idea of \ac{ddos} flows disguised as benign through mimicking the characteristics of normal flows. For instance, \cite{kuppa2019black, aiken2019investigating} show how the attacker can evade a wide range of classifiers after manipulating features such as the number of bytes and packets, inter-arrival time and packet rate in malicious network traffic.
An advantage of using a \ac{gan} over the real benign samples for perturbation is the wide variety of unique values that a \ac{gan} produces.
A well-trained \ac{gan} produces a unique output for each random input, providing a wide range of fake-benign samples. 

The architecture of the generator neural network is based on the deep convolutional generator of \ac{wgan-gp} \cite{gulrajani2017improved} with minimal changes to match the shape of \ac{lucid} samples.
The architecture is modified to have output shape 10x11, as per the shape of \ac{lucid} samples, and 20 as the dimension of the noise source $z$ to keep it smaller than the dimensionality of the output \cite{gantutorial}.
To train this generator, we utilise the discriminator of the same \ac{wgan-gp} and modify its input shape to be 10x11.
The other hyperparameters of both generator and discriminator are unchanged with respect to the original architecture.
Finally, the output layer in \cite{gulrajani2017improved} uses a \textit{tanh} activation function that produces samples with features between [-1, 1]. We rescale the values between [0, 1] in accordance with the normalisation scheme adopted by \ac{lucid}, where all input features are normalised between 0 and 1 before classification.

\begin{algorithm}[t!]
    \small
	\caption{Adversarial dataset generation algorithm}
	\label{alg:adv_dataset_prep}
	\begin{algorithmic}[1]
		\renewcommand{\algorithmicrequire}{\textbf{Input:}}
		\renewcommand{\algorithmicensure}{\textbf{Output:}}
		\Require $T$: Training dataset; $Y$: True labels of $T$; $F$: List of features to perturb;
		\Ensure $T_{adv}$: Adversarial dataset;
		\State $T_{adv} = Copy(T)$; \Comment{Initialise $T_{adv}$}
		\For{$f$ \textbf{in} $F$}
		\State $z = \mathcal{N}(0, 1)$; \Comment{Source of noise}
		\State $T_g = \mathcal{G}(z)$; \Comment{Generate fake-benign samples}
		\State $T_c = Copy(T)$;
		\If{$f$ == flow\_length}
		\State Fill zero-padded rows in $T_c[Y==1]$;
		\Else
		\State $T_c[Y==1, f]$ = $T_g[:, f]$;
		\EndIf
		\State $T_{adv}.append(T_c)$;
		\EndFor
		\State \Return $T_{adv}$
	\end{algorithmic}
\end{algorithm}

\subsection{The perturbation module}
\label{subsec:pert_mod}

The perturbation module takes a dataset with benign and \ac{ddos} samples, and fake-benign samples produced by the generator neural network and outputs the adversarial training dataset.
The perturbed \ac{ddos} samples in the adversarial training dataset are crafted either by replacing features with values seen in benign traffic or by replacing the zero-padded rows with dummy packets.
Both the dummy packets and the substitute values of features are obtained using the generator model.

Algorithm \ref{alg:adv_dataset_prep} shows the process of generating the adversarial training dataset $T_{adv}$.
The inputs of the algorithm are a dataset $T$ with both benign and unperturbed \ac{ddos} samples, the ground truth $Y$ of the dataset and the list of features to perturb $F$, while the output is the adversarial training dataset $T_{adv}$. $F$ is a subset of the 12 features of a sample, 11 packet header features plus \textit{Flow length}, as described in Section  \ref{subsec:lucid}.
Note that each element in $Y$ is a binary label $\in \{0, 1\}$, where 1 denotes a \ac{ddos} sample, and 0 is a benign sample.

In the algorithm, a copy of $T$ initialises $T_{adv}$ (line 1).
To perturb a feature $f \in F$, we start with feeding a source of noise into the generator model $\mathcal{G}(z)$ to generate a set of fake-benign samples $T_g$; then we make a copy of $T$ named $T_c$ (lines 3--5).
For $f$ the \textit{Flow length} feature, the goal is to perturb the number of packets in each \ac{ddos} sample of $T_c$; and thus the zero-padded rows in each sample are filled with packets from the fake samples of $T_g$ (line 7).
For the other features, the value of feature $f$ in each \ac{ddos} sample of $T_c$ is replaced by a value from the fake samples of $T_g$ (line 9).
Then $T_c$ is appended to $T_{adv}$ (line 11).
In this way, $T_{adv}$ is augmented with adversarial \ac{ddos} samples whose features can be observed in benign traffic.

The resulting $T_{adv}$ is unbalanced, with more malicious than benign samples. To obtain a balanced dataset, we duplicate some of the benign samples until we reach an equal number of malicious and benign samples. Of course, the ground truth $Y$ is updated with the labels of the newly added adversarial and benign samples.
It is important to note that, in contrast to the related work \cite{usama2019generative, abou2020investigating}, \ourtool\ decouples the trained model from the generation of the adversarial samples used for training. This is important to prevent the trained model from influencing the strength of the generated adversarial samples \cite{tramer2017ensemble}.




\subsection{Adversarial training}
\label{subsec:adv_train}

We use an adversarial training dataset $T_{adv}$ in order to train the target model to increase its robustness to adversarially perturbed \ac{ddos} attacks.
As $T_{adv}$ contains benign samples as well as perturbed and unperturbed \ac{ddos} samples, the goal of training is to minimise the error between the ground truth and the predictions through optimising the learnable parameters of the model.

Having presented the modules of \ourtool, we revisit Figure \ref{fig:aml_defence_territory} to summarise the end-to-end approach.
The input for the perturbation module is a labelled training dataset $T$, which contains unperturbed \ac{ddos} samples and benign samples.
Then, the generator neural network \edited{receives a number of noise samples from the Gaussian noise source $z = \mathcal{N}(0, 1)$ to generate an equal number of fake-benign samples $T_g$} to perturb the \ac{ddos} samples, as explained in Algorithm \ref{alg:adv_dataset_prep}.
This process produces the adversarial training dataset $T_{adv}$, which is then used to train the model, \ac{lucid}.
The outcome of the \ourtool\ approach is a robust \ac{ddos} detection model capable of distinguishing between benign samples and \ac{ddos} samples even under adversarial attack.

\subsection{Baseline approaches}
\label{subsec:baseline}

\ourtool\ is compared against two baseline approaches that can generate perturbed \ac{ddos} samples for adversarial training.
In the first approach, which we refer to as \ac{bfp}, the training \ac{ddos} samples are perturbed with features obtained from the benign samples in the training dataset.
We use the same procedure explained in Algorithm \ref{alg:adv_dataset_prep} but with $T_g$ from the training data to save the cost of the \ac{gan}.
The second approach is based on the renowned \ac{fgsm} \cite{goodfellow2014explaining} algorithm that uses the gradients of the neural network to create adversarial samples.
We use \ac{fgsm} with $\ell_{\infty}$ bound on the perturbation magnitude to create adversarial samples with minimal perturbations to fool the model.
For both baselines, the augmented dataset $T_{adv}$ contains benign samples, \ac{ddos} samples, and adversarial \ac{ddos} samples with perturbed features as well as perturbed padding rows.

\section{Design of experiments}\label{sec:exp_design}

This section presents the datasets, the process of training different \ac{ml} models, and the design of the experiments used to evaluate the performance of the \ourtool\ approach in normal and adversarial settings.

\begin{table}[t!]
\centering
\caption{Overview of the datasets}
\label{tab:datasets_samples}
\begin{tabular}{@{}llrrr@{}} 
\toprule
 & Dataset & \#Samples & \#Benign & \#DDoS \\ 
\midrule
\multirow{2}{*}{Training} & Custom-SYN & 52698 & 26349 & 26349 \\
 & UNB201X & 265902 & 132746 & 133156 \\ 
\midrule
\multirow{2}{*}{Evaluation} & Scapy-SYN & 18830 & – & 18830 \\
 & CICIDS2017 & 13193 & – & 13193 \\
\bottomrule
\end{tabular}
\end{table}

\subsection{Datasets}

For our evaluation, we consider two types of \ac{ddos} attacks: TCP SYN flood \cite{syn-flood} and HTTP GET flood \cite{get-flood}. Both aim at exhausting the victim's resources by sending large amounts of requests to the victim server, making its services unavailable to the users. 
We use four datasets: Custom-SYN and Scapy-SYN contain TCP SYN flood attacks while the CICIDS2017 \cite{sharafaldin2018toward} and the UNB201X \cite{doriguzzi2020lucid} contain HTTP GET flood attacks.
An overview of these datasets is provided in Table \ref{tab:datasets_samples}.

\begin{table} [t!]
\centering
\caption{Summary of the methodology for training and evaluation of the LUCID models}
\label{tab:train_eval}
\begin{adjustbox}{width=\columnwidth,center}
\begin{tabular}{llll} 
\toprule
\textbf{Model Name} & \begin{tabular}[c]{@{}l@{}}\textbf{Model} \\ \textbf{Detection Goal}\end{tabular} & \textbf{Training} & \textbf{Evaluation} \\ 
\midrule
Model-SYN & SYN flood & \begin{tabular}[c]{@{}l@{}}Custom-SYN \\training dataset\end{tabular} & \begin{tabular}[c]{@{}l@{}}Traffic traces: Scapy-SYN \\ Perturbed Features: \textit{IP Flags; }\\\textit{TCP Len; Flow length} \\ Crafting tool: Scapy\end{tabular} \\ 
\midrule
Model-HTTP & \begin{tabular}[c]{@{}l@{}}HTTP GET \\flood\end{tabular} & \begin{tabular}[c]{@{}l@{}}UNB201X \\training dataset\end{tabular} & \begin{tabular}[c]{@{}l@{}}Traffic traces: CICIDS2017 \\ Perturbed Features: \textit{Time}\\\textit{TCP Len; Flow length} \\ Crafting tool: Scapy\end{tabular} \\
\bottomrule
\end{tabular}
\end{adjustbox}
\end{table}

\subsubsection{\textbf{Custom-SYN dataset}}
A traffic trace recorded in our testbed. It combines legitimate traffic, e.g. web browsing and ssh, and SYN flood attack traffic generated by using the Hping network tool \cite{sanfilippo2006hping}. 

\subsubsection{\textbf{UNB201X dataset}}
This dataset was introduced in \cite{doriguzzi2020lucid} to enable \ac{lucid} to better detect \ac{ddos} attacks irrespective of the training dataset.
UNB201X is a balanced combination of preprocessed traces from ISCX2012 \cite{shiravi2012toward}, CICIDS2017 and CSECIC2018 \cite{sharafaldin2018toward}\footnote{The UNB201X dataset has kindly been made available by the authors of \cite{doriguzzi2020lucid} to enable our evaluation.}.

\subsubsection{\textbf{Scapy-SYN dataset}}
This dataset contains a SYN flood attack crafted by using the Scapy tool \cite{biondi2005scapy}.
Scapy is an open source program to send, capture, dissect and forge network packets.
It comprises seven different traffic traces:  one  with unperturbed SYN attack packets and six traces that contain SYN packets with different perturbed features:
\begin{enumerate}[label=(\roman*)]
	\item \textit{IP flags}: bit \textit{do not fragment} randomly set to 0 or 1.
	\item \textit{TCP length}: SYN packets with random payload content.
	\item \textit{Padding replacement}: the SYN packet is followed by a random number of dummy packets with a random delay.
	\item \textit{SYN Packet Replication}\footnote{The nature of the TCP SYN flood attack with a single SYN packet constituting a flow makes this attack well suited for experimenting with perturbations on the \textit{Flow length} feature.}: the SYN packet is repeated a random number of times with a variable delay.
	\item \textit{IP flags} \& \textit{TCP length} \& \textit{Padding replacement}: combination of these three perturbations.
	\item \textit{IP flags} \& \textit{TCP length} \& \textit{SYN Packet Replication}: combination of these three perturbations.
\end{enumerate}

With regard to \textit{Padding replacement} and \textit{SYN Packet Replication}, these are perturbations to the \textit{Flow length} feature in which the attacker exploits the zero-padded rows in \ac{ddos} samples to fool the \ac{ddos} detection model. 
The attacker's aim is to force the feature extractor to fill them with a random number of additional packets, making the \ac{ddos} samples less dissimilar to the benign samples (in terms of number of packets per flow).
This can easily be achieved by injecting dummy packets into the network with the same IP address and TCP port of the SYN packet, or by replicating the same SYN packet multiple times within a time window. 

\subsubsection{\textbf{CICIDS2017 dataset}}
The CICIDS2017 dataset contains a wide range of normal and malicious network activities, including \ac{ddos}.
We use the timeslot 3.30PM-5.00PM of the CICIDS2017-Fri7PM trace that contains HTTP \ac{ddos} traffic generated by using LOIC \cite{impervaloic}.
We perturb the CICIDS2017 traffic trace with Scapy to introduce two perturbations: random delays between the \ac{ddos} packets and the fragmentation of those packets. 
The delay between packets is reflected in the packet arrival time, which is one of the features used by LUCID (see Table \ref{tab:feat_description}). 
Similarly, packet fragmentation decreases the \textit{TCP Len} of each packet and thus increases the number of packets in the attack flows.
Both perturbations are particularly suitable for the HTTP \ac{ddos} traffic, as each attack flow comprises several packets.

\subsection{Training and evaluating the models}

We produce two separate models.
A model called \textit{Model-SYN} is trained to detect the SYN flood attacks, while a second model called \textit{Model-HTTP} is trained to detect the HTTP GET flood attacks\footnotemark
\footnotetext{This enables comparison with the original \ac{lucid} paper when using the UNB201X dataset.}.
Both models are based on the LUCID code available at \cite{lucid-github}, hence implemented in Python v3.6.8 using the Keras API v2.2.4 on top of Tensorflow 1.13.1 \cite{tensorflow2015-whitepaper} and the ADAM optimization algorithm.
In the following, we present the details of training and evaluating each model, with and without \ourtool. A summary is provided in Table \ref{tab:train_eval}.

\subsubsection{\textbf{Model-SYN}}
First, this model is trained with the Custom-SYN training dataset in order to detect the SYN flood attacks in normal settings, i.e. to classify unperturbed flood attacks.
Using \ourtool, the same model is adversarially trained with the dataset $T_{adv}$ generated by inserting the Custom-SYN training set as input to Algorithm \ref{alg:adv_dataset_prep}.
As $T_{adv}$ contains adversarial samples perturbed using the \ac{gan} neural network, we expect the model to be robust and correctly classify the SYN flood traffic in adversarial settings.
This model is evaluated on the six perturbed traces of the Scapy-SYN dataset to assess the robustness after using \ourtool.

\subsubsection{\textbf{Model-HTTP}}
To detect HTTP GET flood attacks, we train this model with the UNB201X training dataset.
The trained model is used for attack detection in normal settings, i.e. to classify unperturbed flood attacks.
Using \ourtool, the model is retrained with the dataset $T_{adv}$ generated by using the UNB201X training set as input to Algorithm \ref{alg:adv_dataset_prep}.
In both cases $T_{adv}$ is generated based on $F$ that includes the 11 features of \ac{lucid} samples as well as perturbed padding rows.
We evaluate this model on the manipulated CICIDS2017 traffic traces.
This provides an evaluation based on the practical adversary capability of perturbing network traffic.
The results of these experiments are presented in sections \ref{subsec:results_normal_set} and \ref{subsec:results_adv_set}.

In our experiments, the benign samples in the Custom-SYN and UNB201X test datasets are fixed throughout the experiments, while the \ac{ddos} samples in both datasets are replaced by a similar number of perturbed \ac{ddos} samples for the relevant experiment.
In this way, variations in performance are solely attributed to the adversarially perturbed \ac{ddos} samples, and not to a change in benign samples.


\subsection{Evaluation metrics}

We evaluate the classification accuracy using the F1 score, which combines both the recall and the precision and hence provides an overall measure of the model's performance.
The robustness can also be indicated by the \ac{fnr}, which is the number of misclassified \ac{ddos} samples to the total number of \ac{ddos} samples in a test dataset.
These metrics are formally defined as follows:

{\footnotesize
\begin{gather*}
Pr=\frac{TP}{TP+FP}\quad Re=\frac{TP}{TP+FN} \\ FNR=\frac{FN}{FN+TP} \quad F1=2\cdot\frac{Pr\cdot Re}{Pr + Re}
\end{gather*}
}

\noindent where \textit{Pr=Precision}, \textit{Re=Recall}, \textit{F1=F1 Score},  \textit{FNR=False Negative Rate}, \textit{TP=True Positives}, \textit{FP=False Positives},  \textit{FN=False Negatives}.

\section{Experimental Results}\label{sec:results}

This section presents the detection results of the models trained using the baseline approaches and our adversarial training approach.
We employ these models to classify network traces with perturbed \ac{ddos} traffic.
Moreover, it is important to stress that the \ac{gan} is only used to generate adversarial samples for training.
In the following, we use the difference $\Delta$ in the evaluation metric, e.g. F1 score or \ac{fnr}, to show the effect of using \ourtool\ on the robustness of the trained models.
This difference is formally defined as: \Scale[0.9]{$\Delta = Result (After) - Result (Before)$}.

\subsection{Evaluation in normal settings}
\label{subsec:results_normal_set}

First, we want to measure any negative effect of adversarial training with \ourtool\ on LUCID when tested on unperturbed data. Thus, we want to ensure that training with adversarial samples does not cause any degradation to the performance of LUCID. 
In this regard, Table \ref{tab:exp_unperturbed_datasets} shows the detection results of the trained models on unperturbed test datasets before and after using \ourtool. The results of the baseline approaches are omitted from the table due to space limitation.

Before using \ourtool, the LUCID models are capable of detecting \ac{ddos} flows with F1 scores above $99\%$, which aligns with the results already reported in \cite{doriguzzi2020lucid}.
After training with each of the baseline approaches, the models also maintain their detection accuracy with F1 scores above $99\%$ in all datasets.
When trained with \ourtool, the models maintain high accuracy scores, despite a drop in the F1 score of less than $1.5\%$. 
Notably, we observe a slight drop in precision on the SYN datasets, and in recall score on the CICIDS2017 dataset.
This indicates that a few ambiguous network flows are misclassified, which matches the observations made in \cite{zhang2020tiki} after adversarial training and \cite{apruzzese2020appcon} after employing ensemble learning to improve robustness.

\begin{table}[t!]
	\centering
	\caption{Evaluation of Model-SYN and Model-HTTP against unperturbed test datasets before and after using \ourtool}
	\label{tab:exp_unperturbed_datasets}
	\begin{adjustbox}{width=0.9\linewidth,center}
	\begin{tabular}{@{}lcccccc@{}}
	\toprule
		\multirow{2}{*}{\textbf{\begin{tabular}[c]{@{}l@{}}Dataset\\ (Model Name)\end{tabular}}} & \multicolumn{3}{c}{\textbf{\begin{tabular}[c]{@{}c@{}}Scapy-SYN\\ (Model-SYN)\end{tabular}}} & \multicolumn{3}{c}{\textbf{\begin{tabular}[c]{@{}c@{}}CICIDS2017\\ (Model-HTTP)\end{tabular}}} \\ \cmidrule(l){2-7}
	 & \textbf{Before} & \textbf{After} & \textbf{$\Delta$} & \textbf{Before} & \textbf{After} & \textbf{$\Delta$} \\ \midrule
	\textbf{Precision} & 0.9985 & 0.9739 & -0.0246 & 0.9873 & 0.9902 & 0.0029 \\ \midrule
	\textbf{Recall} & 1.0000 & 1.0000 & 0.0000 & 0.9990 & 0.9978 & -0.0012 \\ \midrule
	\textbf{F1 score} & 0.9992 & 0.9867 & -0.0125 & 0.9931 & 0.9940 & 0.0009 \\ \midrule
	\textbf{FNR} & 0.0000 & 0.0000 & 0.0000 & 0.0009 & 0.0021 & 0.0012 \\ \bottomrule
	\end{tabular}
	\end{adjustbox}
\end{table}

\subsection{Evaluation in adversarial settings}
\label{subsec:results_adv_set}

\begin{table*}[t!]
\centering
\caption{Evaluation of Model-SYN against perturbed traces of the Scapy-SYN dataset}
\label{tab:pert_syn_trace_results}
    \begin{adjustbox}{width=0.9\textwidth,center}
        \begin{tabular}{@{}lll|llll|llll|llll@{}} \toprule
            \multirow{2}{*}{\textbf{Perturbations}} & \multicolumn{2}{c}{\textbf{Before}} & \multicolumn{4}{c}{\textbf{\ourtool}} & \multicolumn{4}{c}{\textbf{\ac{bfp}}} & \multicolumn{4}{c}{\textbf{FGSM}} \\ \cmidrule{2-15}
             & \textbf{F1 Score} & \textbf{FNR} & \textbf{F1 Score} & \textbf{$\Delta$} & \textbf{FNR} & \textbf{$\Delta$} & \textbf{F1 Score} & \textbf{$\Delta$} &\textbf{FNR} & \textbf{$\Delta$} & \textbf{F1 Score} & \textbf{$\Delta$} &\textbf{FNR} & \textbf{$\Delta$} \\ \midrule
            \textbf{IP Flags} & \textbf{0.6596 } & \textbf{0.5071 } & \textbf{0.9867 } & \textbf{0.3271} & \textbf{0.0003 } & \textbf{-0.5068} & 0.6558 & -0.0038 & 0.5071 & 0.0000 & 0.6541 & -0.0055 & 0.5071 & 0.0000 \\ \midrule
            \textbf{TCP Len} & 0.9992 & 0.0000 & 0.9867 & -0.0125 & 0.0000 & 0.0000 & 0.9953 & -0.0039 & 0.0000 & 0.0000 & 0.9933 & -0.0059 & 0.0000 & 0.0000 \\ \midrule
            \textbf{SYN Packet Replication} & 0.9547 & 0.0851 & 0.9867 & 0.0320 & 0.0000 & -0.0851 & 0.9951 & 0.0404 & 0.0000 & -0.0851 & 0.9933 & 0.0386 & 0.0000 & -0.0851 \\ \midrule
            \textbf{Padding Replacement} & \textbf{0.7875 } & \textbf{0.3494 } & \textbf{0.9867 } & \textbf{0.1992} & \textbf{0.0000 } & \textbf{-0.3494} & \textbf{0.9757} & \textbf{0.1882} & \textbf{0.0382} & \textbf{0.3112} & 0.9127 & 0.1252 & 0.1491 & -0.2003 \\ \midrule
            \begin{tabular}[c]{@{}l@{}}\textbf{IP Flags; TCP Len; SYN}\\\textbf{ Packet Replication}\end{tabular} & \textbf{0.6322 } & \textbf{0.5368 } & \textbf{0.9846 } & \textbf{0.3524} & \textbf{0.0041 } & \textbf{-0.5327} & \textbf{0.9888} & \textbf{0.3566} & \textbf{0.0130} & \textbf{-0.5238} & 0.9066 & 0.2744 & 0.1596 & -0.3772 \\ \midrule
            \begin{tabular}[c]{@{}l@{}}\textbf{IP Flags; TCP Len;}\\\textbf{ Padding Replacement}\end{tabular} & \textbf{0.5862 } & \textbf{0.5843 } & \textbf{0.9867 } & \textbf{0.4005} & \textbf{0.0000 } & \textbf{-0.5843} & \textbf{0.9731} & \textbf{0.3869} & \textbf{0.0431} & \textbf{-0.5412} & 0.5999 & 0.0137 & 0.5657 & -0.0186 \\ \bottomrule
        \end{tabular}
    \end{adjustbox}
\end{table*}

In what follows, we test Model-SYN and Model-HTTP on perturbed flood attacks that can be generated according to the attacker's knowledge as defined in our threat model (Section \ref{sec:threat_model}).
Moreover, we demonstrate the positive impact of using \ourtool\ on the models' robustness.

\subsubsection{\textbf{Model-SYN}}
The F1 scores of Model-SYN before and after using \ourtool\ for training, and tested on perturbed samples, are shown in Table \ref{tab:pert_syn_trace_results}.

With no adversarial training, we observe a substantial sensitivity of the model to adversarial SYN flood attacks built with perturbations on \textit{IP flags}, and \textit{Padding replacement}, which leads to a drop in the F1 score to $65.96\%$, and $78.75\%$, respectively.
The results obtained with \textit{IP flags} confirm the high ranking of this feature in the analysis presented in the original LUCID paper \cite{doriguzzi2020lucid}. On the other hand, the impact of manipulation of the \textit{Flow length} feature through \textit{Padding replacement} can be explained by analysing the properties of the SYN flood attack traffic. Indeed, the \textit{Flow length} of the SYN samples is either one or two packets (the SYN packet of the attacker and the SYN-ACK from the victim server, if its connection tables are not full), while benign samples in our datasets are longer on average. This difference becomes less prominent when the \textit{Flow length} feature  of \ac{ddos} samples are perturbed  with \textit{Padding replacement}, leading to a decrease in the classification accuracy.

These results demonstrate that perturbing the \textit{IP flags} and replacing the padding rows with dummy packets can generate a successful evasion attack.
We also identify that similar success can be achieved by perturbing a combination of features in the same SYN network flow.
For instance, while introducing \textit{SYN Packet Replication} perturbation has decreased the F1 score to $96 \%$, combining it with the \textit{IP flags} and \textit{TCP length} perturbations yields a further $33 \%$ F1 score decrease, producing a stronger evasion attack.

We also observe that adversarial training might produce a decrease in the classification accuracy when perturbed samples are present. This is indicated by the negative $\Delta$ values in Table \ref{tab:pert_syn_trace_results}. Although minimal, the decrease is caused by the change in the distribution of the features due to the adversarial samples added to the training set. This change moves the decision boundary, and hence the accuracy scores, F1 included.

Adversarial training with the \ac{bfp} baseline approach allows the Model-SYN to restore most of its detection accuracy with the majority of F1 scores $>97\%$ except for the \textit{IP Flags} perturbation.
This indicates that adversarial training with this approach leaves the model vulnerable to perturbation of features not present in the benign samples.
The \ac{fgsm}-based approach leads to a minor improvement in robustness against most of the perturbations. 
This minor increase in robustness suggests that \ac{fgsm} or similar methods (e.g., \ac{pgd}) might not be the best option to generate adversarial examples for the training of \ac{nids}.
On the other hand, after retraining using \ourtool, we obtain an overall higher classification accuracy $(>98\%)$ and a substantial reduction in the false negative rate, which drops to reach approximately $0\%$ for all perturbations. 
The results highlight that the adversarial dataset $T_{adv}$ created with \ourtool\ covers the space of adversarial perturbations that attackers might exploit.
This also indicates that Model-SYN is more robust to adversarial attacks with adversarial training using \ourtool\ than without.

\subsubsection{\textbf{Model-HTTP}}

\begin{table*}[t!]
\centering
\caption{Evaluation of Model-HTTP against perturbed traces of the CICIDS2017 dataset}
\label{tab:pert_2017_trace_results}
\begin{adjustbox}{width=0.9\textwidth,center}
\begin{tabular}{@{}lll|llll|llll|llll@{}} \toprule
\multirow{2}{*}{\textbf{Perturbations}} & \multicolumn{2}{c}{\textbf{Before}} & \multicolumn{4}{c}{\textbf{\ourtool}} & \multicolumn{4}{c}{\textbf{\ac{bfp}}} & \multicolumn{4}{c}{\textbf{FGSM}} \\ \cmidrule(l){2-15}
 & \textbf{F1 Score} & \textbf{FNR} & \textbf{F1 Score} & \textbf{$\Delta$} & \textbf{FNR} & \textbf{$\Delta$} & \textbf{F1 Score} & \textbf{$\Delta$} & \textbf{FNR} & \textbf{$\Delta$} & \textbf{F1 Score} & \textbf{$\Delta$} & \textbf{FNR} & \textbf{$\Delta$} \\ \midrule
\textbf{Delay} & \textbf{0.5516} & \textbf{0.6150} & \textbf{0.9871} & \textbf{0.4355} & \textbf{0.0177} & \textbf{-0.5973} & 0.9038 & 0.3522 & 0.1659 & -0.4491 & 0.8105 & 0.2589 & 0.3124 & -0.3026 \\ \midrule
\textbf{Packet Fragmentation} & 0.9406 & 0.1024 & 0.9935 & 0.0529 & 0.0000 & -0.1024 & 0.9879 & 0.0473 & 0.0112 & -0.0912 & 0.8833 & -0.0573 & 0.2019 & 0.0995 \\ \bottomrule
\end{tabular}
\end{adjustbox}
\end{table*}

As noted in Section \ref{subsec:lucid}, packet arrival \textit{time} and \textit{TCP Len} are important features in the detection model. 
We therefore measure the impact of delay perturbation and packet fragmentation on the detection accuracy.
Packet fragmentation perturbs the \textit{TCP Len} feature of each packet and increases the number of packets in each flow as well. 
Indeed, packet arrival time, packet size and the number of packets in a flow are features employed to generate adversarial network flows across the state-of-the-art, e.g., \cite{zhang2020tiki, hashemi2019towards, hashemi2020enhancing, aiken2019investigating, yan2019automatically}. 
We apply the perturbations to the \ac{ddos} packets of the HTTP-GET flows in the CICIDS2017 dataset. 
As shown in Table \ref{tab:pert_2017_trace_results}, with no adversarial training, the HTTP GET attack with a delay perturbation generates a strong evasion attack, as the \ac{fnr} increases to $61.5\%$, confirming the importance of this feature for the detection model.   
On the contrary, packet fragmentation generates a weaker evasion attack with \ac{fnr} around $10\%$.
The reason is that the attack flows naturally comprise several packets, and hence packet fragmentation has a less noticeable impact on the attack flows. 

With adversarial training using the baseline approaches, it is evident that the trained models remain vulnerable to \ac{ddos} attacks with perturbed samples. The better of the two models, which is the \ac{bfp} approach, is still vulnerable to the delay perturbation as the F1 score and \ac{fnr} are approximately $90\%$ and $16\%$ respectively. 
After using \ourtool\ for training, the detection accuracy of the model in the case of delay perturbation experiences more than $43\%$ improvement in the F1 score, reaching $98.71\%$. The same improvement is also apparent in the \ac{fnr} that drops to $1.8\%$, indicating the increased resilience to adversarial perturbations.
A similar level of resilience is also noticed in the case of the packet fragmentation as the \ac{fnr} has dropped to $0\%$.
This highlights that, in comparison with the baseline approaches, \ourtool\ is able to create adversarial training dataset $T_{adv}$ with a wide coverage of adversarial perturbations. 
These results, combined with those of Model-SYN, demonstrate the considerable improvement in robustness after using \ourtool.

\subsection{Discussion}

The results presented in the previous sections confirm that \ac{ml}-based \acp{nids} are vulnerable to adversarial attacks.
Potential adversaries can exploit their domain expertise to craft perturbed flood attacks without requiring knowledge of the underlying detection model.
For example, they can perturb the delay in the network flows of the HTTP GET flood attack, as we have done in the CICIDS2017 dataset, which causes the \ac{fnr} (misclassified \ac{ddos} samples) to reach $61.5 \%$.
Building upon this \ac{fnr}, they can increase the number of bots (botnet \ac{ddos} attacks) to exhaust the victim resources and achieve a successful \ac{ddos} attack.
In contrast, after adopting \ourtool, most of the perturbed flood samples are detected, with the \ac{fnr} reduced to below $1.8 \%$.

The models trained using \ourtool\ achieve high performance with F1 scores above $98 \%$ on perturbed \ac{ddos} network traces, improving on the $62 \%$ average detection rate reported in \cite{hashemi2020enhancing}.
We highlight that the models trained using \ourtool\ have detected the SYN and HTTP GET flood attacks, whether they are perturbed or not, allowing for the mitigation of such attacks.
\edited{Moreover, the architecture of those models remains unchanged, and thus \ourtool\ has no impact on their performance in terms of processed samples/second and memory requirements.}

Finally, a consideration for practical deployment of a ML system such as \ourtool\ is prediction interpretability \cite{jacovi2018understanding}. We undertook a simple study of how our trained model classifies samples after adversarial training by exploring the filter activations of a convolution layer while processing samples. This revealed a change in the filter activation across different features as well as a shift in their ranking, and hence their contribution to the model's decision. For example, our initial analysis indicated that there is no longer a heavy dependence on a single feature such as \textit{Highest layer}, as reported in \cite{doriguzzi2020lucid}, which suggests to us that the adversarial training evens the activation across multiple features, which results in the increased robustness to evasion attacks. However, this is not conclusive and requires an in-depth analysis, which we propose for our future work.

\section{Conclusion}\label{sec:conclusion}

In this paper, we introduced \ourtool, an approach for adversarial training of \ac{ml} models detecting \ac{ddos} flood attacks.
We trained two \ac{lucid} models, which are state-of-the-art for \ac{ddos} detection, to detect two types of flood attacks, namely, SYN flood and HTTP GET flood.
However, these models, as is the case for other \ac{ml} models, are vulnerable to adversarially perturbed \ac{ddos} samples and traces that aim to evade such models, reaching the victim machine and achieving the DoS goal.
To defend against such adversarial attacks, we trained both models using \ourtool, which leverages a \ac{gan} to generate adversarial \ac{ddos} samples for training.
The trained models were evaluated on real perturbed traces generated without requiring knowledge of the victim model. This approach sets \ourtool\ apart from other similar solutions.
The results show that the models trained with \ourtool\ can detect with high accuracy both the SYN and HTTP flood attacks, regardless of whether they are perturbed or not.
The models achieved F1 score above $98 \%$ and the \ac{fnr} drops to below $1.8 \%$ on perturbed \ac{ddos} network traces.
Future work will include extending \ourtool\ to other \ac{ddos} attack types and studying the use of open source tools, such as LOIC, to generate perturbed \ac{ddos} attacks.

\bibliography{bibliography}
\bibliographystyle{IEEEtran}

\end{document}